# On the Depolarization Factors for Anisotropic Ellipsoids in Anisotropic Media

L. A. Apresyan,  D. V. Vlasov,

Prokhorov General Physics Institute, Russian Academy of Science

ul.Vavilova 38, Moscow, 119991 Russia

**Abstract**

A definition of depolarization factors for anisotropic ellipsoid in an anisotropic medium is considered. The expressions for these factors are derived, which generalize some results known for special cases, and differences from the usual depolarization factors for ellipsoid in isotropic medium are described. The concept of "reduced" ellipsoid, allowing visually describe the effect of the medium anisotropy on depolarization is also introduced.

### 1.Introduction

The solution of classical problem of homogeneous isotropic ellipsoid in a uniform external field in an isotropic medium is described in many textbooks and used extensively in applications (see, eg., [1,2]; we call this case for the sake of brevity, "isotropic"). "Second Wind" to usage this solution is related with the development of nanotechnology, and in particular, nanocomposites, where the model of a homogeneous ellipsoid is now widely used, including the description of nanoparticles, such as carbon nanotubes and graphene [3].  Less well known is the description of general anisotropic case, namely, the anisotropic ellipsoid placed in a uniform external field in an anisotropic medium. For this case the explicit form of exact solution is also known [4].

In this methodological paper we consider the relation between isotropic and anisotropic cases, introducing the depolarization factors for anisotropic problem, and also the notion of "reduced ellipsoid", that allows us to express a general view on the depolarization factors in a particularly simple manner. The use of these factors makes it easier to trace the differences between anisotropic and isotropic cases, avoiding inaccuracy that sometimes allowed in the literature, when the depolarization factor for isotropic problem misused in anisotropic case (see, eg., [7,8]).

In the case of a spheroidal particle coaxial with a uniaxial anisotropic medium, for the depolarization factors some analytical expressions are obtained generalizing the results known from the literature. In the last case the depolarization factors depend on the ratio of square of the aspect ratio of the ellipsoid to the factor of anisotropy of the medium, which is defined as the ratio of distinct eigenvalues of the dielectric permittivity tensor, or for another statement, of static conductivity tensor of anisotropic medium.

## 2. Definition of depolarization factors for ellipsoid in an anisotropic medium

In order to define the concept of depolarization factors for anisotropic ellipsoid in an anisotropic medium, we use the well-known solution of this problem for the case of a uniform external field [4,8] (simplified derivation of this solution is given in Appendix A). For definiteness, we consider a dielectric ellipsoid with an anisotropic dielectric constant, although the final results can be easily applied to the case of a conductive ellipsoid, as well as similar problems in the theory of magnetism, heat conduction and diffusion. At the same time, to avoid cluttering the formulas, we will not introduce special symbols for tensor and vector quantities, considering their nature sufficiently clear from the context[1]

In the case of a uniform external electric field $E^\infty$ the field E inside ellipsoid is uniform, $E = E^{in} = \text{const}$ and is expressed through an external field by the relation (see Appendix, (A10))

$$E^{in} = (1 - \Gamma \delta\varepsilon)^{-1} E^\infty. \qquad (1)$$

Here $\delta\varepsilon = \varepsilon_i - \varepsilon_0$, $\varepsilon_i$ and $\varepsilon_0$ - dielectric tensor of ellipsoid and medium, respectively, matrix $\Gamma$ is defined by integral over volume V of ellipsoid

$$\Gamma = - \int_V \nabla\nabla G_r \, dr, \qquad (2)$$

where $G_r$ - scalar Green's function of a homogeneous anisotropic medium (see Appendix, (A4)), and gradients constitute tensor product, $(\nabla\nabla)_{ij} := \nabla_i \nabla_j$. From (2) it is seen that the matrix $\Gamma$ is symmetric, ie does not change when transposing: $\Gamma = \Gamma^T$, or in components, $\Gamma_{ij} = \Gamma_{ji}$.

Tensor $\Gamma$ was introduced in [7] in the form of following from (2) surface integral

$$\Gamma = - \oint_S \nabla G_r \, n \, d^2r, \qquad (3)$$

where S - the surface of ellipsoid and n - unit outward normal to the surface (so $n \, d^2r$ is vector element of S).

In order to link the tensor $\Gamma$ with conventional depolarization factors $L_j$, which are defined for isotropic ellipsoid in an isotropic medium (for which $\varepsilon_0$ and $\delta\varepsilon$ scalar), we compare the expression (1) with similar expression for components $E_j^{in}$ for ellipsoid in the case of isotropic medium [1, 2]:

$$E_j^{in} = (1 + L_j \frac{1}{\varepsilon_0} \delta\varepsilon)^{-1} E_j^\infty \qquad (4)$$

or, in symmetric tensor form,

$$E^{in} = (1 + \frac{1}{\sqrt{\varepsilon_0}} L \frac{1}{\sqrt{\varepsilon_0}} \delta\varepsilon)^{-1} E^\infty \qquad (5)$$

where tensor L is diagonal in the principal axes of the ellipsoid with eigenvalues $L_j$.

Comparing (5) to (1) shows that for generalization on anisotropic case it is reasonable to define the geometric factor - tensor L by the relation

---

[1]When using the bellow expressions it is convenient to treat most of the values a priori as non-commuting tensors, considering the emergence of scalars as smiles of fortune.

$$\Gamma = - \frac{1}{\sqrt{\varepsilon_0}} L \frac{1}{\sqrt{\varepsilon_0}}, \qquad (6)$$

so that

$$L = - \sqrt{\varepsilon_0} \, \Gamma \, \sqrt{\varepsilon_0}. \qquad (7)$$

Taking into account, that tensor $\varepsilon_0$ is symmetric, $\varepsilon_0^T = \varepsilon_0$, such generalization to the case of anisotropic $\varepsilon_0$ preserves symmetry L, allowing to determine the geometrical factors as eigenvalues of L. In the case of an isotropic medium, when tensor $\varepsilon_0$ reduces to a scalar, (6) simplifies and gives

$$L = - \varepsilon_0 \, \Gamma. \qquad (8)$$

In this problem when calculating $\Gamma$ (3) there are two sources of anisotropy: the anisotropy of dielectric tensor $\varepsilon_0$ and shape anisotropy of the ellipsoid. The last can be described by "axes tensor "A, which is diagonal in the basis of the principal axes of the ellipsoid: in this basis A = diag($a_1$, $a_2$, $a_3$), where $a_i$ - semiaxes of the ellipsoid. Then the equation of the ellipsoid surface S can be written as

$$r^T A^{-2} r \equiv \sum_{i=1,2,3} \frac{r_i^2}{a_i^2} = 1, \qquad (9)$$

where the superscript "T" denotes transposition. For brevity call ellipsoid (9) «A-ellipsoid».

Substituting the expression for $\Gamma$ (3) into (7), after the change of variables $r \Rightarrow \sqrt{\varepsilon_0} \, r$ and transition from integral over surface to the integration over solid angle element $d\Omega_r$ near the direction of unit vector $\hat{r} = r / |r|$, $dS = \hat{n} \, r^2 d\Omega_r / (\hat{r}^T \cdot \hat{n})$, for the depolarization tensor L we obtain a simple expression

$$L = < \frac{\hat{r} \, \hat{n}}{(\hat{r}^T \cdot \hat{n})} > \equiv \int \frac{\hat{r} \, \hat{n}}{(\hat{r}^T \cdot \hat{n})} \frac{d\Omega_r}{4\pi}, \qquad (10)$$

which is the main result of this work. Here dyad $\hat{r} \, \hat{n}$ is tensor product (ie, second rank tensor with components $r_i \, n_j$), $(\hat{r}^T \cdot \hat{n})$ - scalar product and $\hat{n}$ - the normal to the "reduced" ellipsoid, whose surface is described by the equation:

$$r^T \cdot A_\varepsilon^{-2} \cdot r \equiv r^T \cdot \sqrt{\varepsilon_0} \, A^{-2} \sqrt{\varepsilon_0} \cdot r = 1, \qquad (11)$$

(it is obvious, that instead of unit vectors $\hat{r}$ and $\hat{n}$ we can use in (10) arbitrary normalized vectors with directions of $\hat{r}$ and $\hat{n}$).

From (11) it follows that the "reduced" ellipsoid corresponds to symmetric axes tensor $A_\varepsilon$, defined by the relation

$$A_\varepsilon^2 = \frac{1}{\sqrt{\varepsilon_0}} A^2 \frac{1}{\sqrt{\varepsilon_0}}. \qquad (12)$$

The directions of principal axes of $A_\varepsilon$ do not coincide with the directions of the principal axes of the ellipsoid (tensor A), or with the directions of the principal axes of the medium (tensor $\varepsilon_0$ ), but are expressed only as a combination of both. Thus, the "reduced" $A_\varepsilon$ tensor takes into account the anisotropy of ellipsoid and the anisotropy of medium, and the geometrical factor L depends only on $A_\varepsilon$ (8), L = L ($A_\varepsilon$ ), but not from $\varepsilon_0$ and A separately.

### 3. Some consequences

Consider some simple consequences of (10). Since the trace of a dyad is equal to scalar product, Sp $\hat{r}$ $\hat{n}$ = ($\hat{r}^T \cdot \hat{n}$ ), from (10) it is immediately clear that, as in the isotropic case, the trace of L is equal to one:

$$\text{Sp L} = \left< \frac{(\hat{r}^T \cdot \hat{n})}{(\hat{r}^T \cdot \hat{n})} \right> = 1. \qquad (13)$$

This fact can be used to simplify calculations (note that the original tensor $\Gamma$ has no such property).

The expression (10) allows us to give a simple geometric sense to the tensor L. Indeed, according to (10) L can be interpreted as normalized by (13) tensor correlation of the unit direction vector $\hat{r}$ with the corresponding unit normal $\hat{n}$, obtained by averaging over all possible directions of $\hat{r}$ (or in other words, on the surface of the "reduced" ellipsoid $A_\varepsilon$ ) in accordance with (10).

If we consider the eigenvalues of $A_\varepsilon$ and direction of its own axis $e_i$ as known, so that we have a canonical representation

$$A_\varepsilon = \Sigma\, a_{\varepsilon\, i}\, e_i\, e_i, \qquad (14)$$

tensor L in the basis $e_i$ is diagonal with eigenvalues

$$L_i = \left< \frac{\hat{r}_i\, \hat{n}_i}{(\hat{r}^T \cdot \hat{n})} \right>, \qquad (15)$$

where $\hat{r}_i$ and $\hat{n}_i$ are i-component of $\hat{r}$ and $\hat{n}$, respectively. The relation (15) generalizes to the anisotropic case usual depolarization factors for ellipsoid in isotropic medium.

In the case of isotropic medium, when $\varepsilon_0$ reduces to scalar, from (12) we have

$$A_\varepsilon = \frac{1}{\sqrt{\varepsilon_0}}\, A, \qquad (16)$$

In this case, the principal axes of the tensors $A_\varepsilon$ and A are the same, and the eigenvalues differ only by a factor, so that they define similar ellipsoids. As the normals to such ellipsoids for each direction $\hat{r}$ are the same, upon calculation of (8) one can use «reduced» $A_\varepsilon$-ellipsoid instead of the original A-ellipsoid. The canonical representation (14) for A-ellipsoid can be regarded as known (eigenvectors A directed along principal axes of ellipsoid), so that the geometrical factors (15) are determined only by the geometry of ellipsoid and not depend on dielectric constant of medium.

As an example of using (10), consider the case of coaxial ellipsoid and medium, which is next by complexity after the isotropic case. In this case in (14) the eigenvectors $e_i$ are directed along the principal axes of ellipsoid, and the eigenvalues from (12) are $a_{\varepsilon\, i} = \frac{a_i}{\sqrt{\varepsilon_i}}$. We restrict ourselves to the assumption that the first two eigenvalues are the same, so that the "reduced" ellipsoid is spheroid, and all the integrals can be taken explicitly.

Then, choosing (x,y,z)-basis with z axis directed along the third principal axis of ellipsoid and calculating from the equation for "reduced" ellipsoid (11) the components of normal vector $\hat{n}$ (which is proportional to gradient of the left side of (11)), for eigenvalue $L_3 = L_z$ after transition to the spherical coordinates $(\theta, \varphi)$ and integrating over the azimuthal angle $\varphi$ from (15) we have

$$L_z = \left< \frac{z n_z}{(x n_x + y n_y + z n_z)} \right> = \left< \frac{g \cos^2 \theta}{\sin^2 \theta + g \cos^2 \theta} \right>$$

$$= \frac{g}{2} \int_{-1}^{1} \frac{\mu^2}{1-(1-g)\mu^2} d\mu = \frac{g}{1-g} \left( \frac{\operatorname{arcth}\sqrt{1-g}}{\sqrt{1-g}} - 1 \right). \quad (17)$$

Here $\mu = \cos \theta$, $g = v_a^2 / v_\varepsilon$ - the ratio of two factors anisotropy: ellipsoid $v_a^2 = (a_\perp/a_z)^2$ and environment $v_\varepsilon = \varepsilon_\perp / \varepsilon_z$ (to be specific in evaluating the integral (17) we get $g < 1$). The remaining eigenvalues $L_{1,2}$ can be expressed using the normalization condition (13), $L_1 + L_2 + L_3 = 1$, which gives $L_1 = L_2 = (1 - L_z)/2$. Note that in numerical calculations it can be more convenient to use directly integral (17), which is defined for all values of g.

In the case of an isotropic medium $v_\varepsilon = 1$, $g = v_a^2$ and (17) reduces to the known expression for the geometric factor of prolate ellipsoid in an isotropic medium [5,6]. In the case of the sphere in an anisotropic medium $v_a = 1$, so that $g = 1/v_\varepsilon$. In this case, taking into account following from (6) relation

$$\Gamma_z = -\frac{1}{\varepsilon_z} L_z, \quad (18)$$

and calculating the integral (17) for $g > 1$, we have

$$\Gamma_z = -\frac{1}{\varepsilon_z} \frac{g}{g-1} \left( 1 - \frac{\operatorname{atctg}\sqrt{g-1}}{\sqrt{g-1}} \right). \quad (19)$$

This expression is equivalent to the analogous one obtained in [7]. Diagonal components $\Gamma_{1,2}$ in analogy with (18) can be directly expressed in terms of the components $L_{1,2} = (1 - L_z)/2$. Thus, (17) allows us to extend the known results [1,7] to the anisotropic case.

It follows from (17) that in the case of co-axial ellipsoid and medium the geometrical factors $L_j$ depend only on the ratio $g = v_a^2 / v_\varepsilon$ of ellipsoid factor anisotropy $v_a^2$ and environment factor $v_\varepsilon$, which differences from unity characterize the appropriate degree of anisotropy. The dependence $L_j(g)$ reflects the fact that for depolarization in an anisotropic medium essential is not only the geometrical shape of ellipsoid, but first of all the form of "reduced" ellipsoid (11). In particular, the ellipsoid can be oblate and have nevertheless depolarization factor, which in isotropic medium would be characteristic for a prolate ellipsoid - it is enough for this that "reduced" ellipsoid be prolate.

In the literature one often uses the value of aspect ratio $\alpha = a_z/a_\perp = 1/v_a$, so that for isotropic problem for disks $\alpha \ll 1$, $L_z \to 1$, for sphere $\alpha = 1$, $L_z = 1/3$, and for needles $\alpha \gg 1$, $L_z \to 0$. To illustrate the effect of the medium anisotropy on $L_z$ the plots $L_z(\alpha)$ (17) are shown in Fig.1 for different values $v_\varepsilon$. These plots show that though the character of the $L_z(\alpha)$ - dependence in anisotropic medium is stored, for $\nu_\varepsilon \neq 1$ the effect of anisotropy on the depolarization is significant, and this should be taken into account in applications where such depolarization is important.

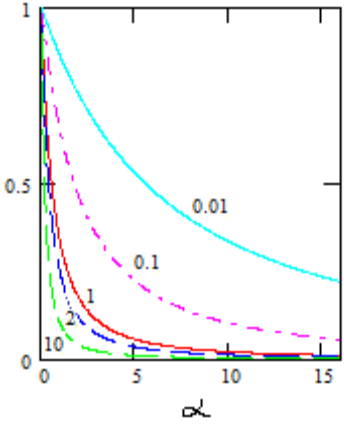

Fig.1. The dependence of depolarization factor $L_z$ from aspect ratio at different levels of anisotropy degree $\nu_\varepsilon$ ($\nu_\varepsilon$ values are shown near the curves)

## 4. Conclusion

In this paper we have considered generalization of ellipsoid depolarization factors, usually defined only for the case of isotropic media, to the case of anisotropic media with tensor dielectric characteristics. The use of such factors can simplify calculations, and in addition, help better understand the structure of solutions and to avoid some errors due to the misuse of the depolarization factors of isotropic problem in general anisotropic case.

**Appendix A: Anisotropic problem for ellipsoid in an external field**

Consider the problem of homogeneous ellipsoid with dielectric permittivity tensor $\varepsilon_i$ in homogeneous anisotropic medium with dielectric tensor $\varepsilon_0$, placed in a uniform external electric field $E^\infty$. It is required to determine the distribution of the field E, both inside and outside ellipsoid. The solution of this problem is known [7, 4]. For convenience, we repeat basically calculations [7], giving a simplified derivation, which allows to write the solution used in the main text. This approach is much easier then rather complicated reasoning [5] which used the analogy with the theory of elasticity and Eshelby tensors (note that in [5] instead of considered here tensor $\Gamma$ asymmetric tensor $S = -\varepsilon_0 \Gamma$ was used).

Introduce the characteristic function of ellipsoid $\theta_v(r)$: $\theta_v(r) = 1$ inside and $\theta_v(r) = 0$ outside ellipsoid. Then the distribution of dielectric permittivity can be written as

$$\varepsilon(r) = \varepsilon_0 (1 - \theta_v(r)) + \varepsilon_i \theta_v(r) = \varepsilon_0 + \delta\varepsilon(r), \qquad (A1)$$

where $\delta\varepsilon(r) = \delta\varepsilon\, \theta_v(r)$, $\delta\varepsilon = \varepsilon_i - \varepsilon_0$, and both $\varepsilon_i$ and $\varepsilon_0$ are considered to be symmetrical. Bearing in mind that the field E in static problem is potential, ie expressed as the gradient of the potential $\Phi$, $E = - \nabla \Phi$, from the constitutive equation $D = \varepsilon E$ from the Maxwell equation in the absence of external charges $\nabla D = 0$, we have equation for $\Phi$:

$$\nabla \varepsilon (r) \nabla \Phi = 0, \qquad (A2)$$

or, using (A1)

$$\nabla \varepsilon_0 \nabla \Phi = - \nabla \delta\varepsilon (r) \nabla \Phi, \qquad (A2)`$$

supplemented with boundary condition

$$\Phi\big|_{S^\infty} = - rE^\infty \equiv \Phi^\infty (r) \qquad (A3)$$

at the infinite radius sphere $S^\infty$.

To obtain the integral form of (A2), (A3), consider the Green's function of a homogeneous anisotropic medium

$$G_r = - \frac{1}{4\pi\sqrt{|\varepsilon_0|\left(r \cdot \frac{1}{\varepsilon_0} \cdot r\right)}}, \qquad (A4)$$

which satisfies the equation

$$\nabla \cdot \varepsilon_0 \cdot \nabla G_r = \delta(r), \qquad (A5)$$

corresponding to the operator on the left side (A2)` (in (A4) $|\varepsilon_0|$ = det $\varepsilon_0$ is determinant $\varepsilon_0$, and the point is convolution for neighboring indices).

With this function, the problem (2), (3) can be written as

$$\Phi = \Phi^\infty - \int G_{r-r`} \nabla` \delta\varepsilon (r`) \nabla` \Phi(r`) dr`$$

$$= \Phi^\infty + \int_V (\nabla` G_{r-r`}) \delta\varepsilon E(r`) dr`, \qquad (A6)$$

where integral by parts was used. Taking the gradient of both sides of (6), we obtain the equation for E

$$E(r) = E^\infty + \int_V (\nabla \nabla` G_{r-r`}) \delta\varepsilon E(r`) dr`$$

$$= E^\infty + \int_V (\nabla \nabla` G_{r-r`}) dr` \delta\varepsilon E_{in}, \qquad (A7)$$

In the last integral we have taken into account that in the case of an ellipsoid the field E inside it is constant, $E(r) = E_{in}$ (see [7,8]). Setting in (A7) $r \in V$, we obtain closed matrix equation for the constant field $E_{in}$

$$E_{in} = E^\infty + \Gamma \delta\varepsilon E_{in}, \qquad (A8)$$

where

$$\Gamma = \int_V \nabla \nabla` G_{r-r`} \, dr` = - \int_V \nabla` \nabla` G_r \, dr`. \qquad (A9)$$

Here we take into account that for $r \in V$ the integral in (A9) is independent from $r$, so that it is convenient to put $r$ in center of ellipsoid.

Solving the matrix equations (8), for the field inside ellipsoid $E_{in}$ we find

$$E_{in} = (1 - \Gamma \delta\varepsilon)^{-1} E^\infty. \qquad (A10)$$

Substituting this expression into the right side of (A7), we obtain the field E for all values of r, and in particular, outside ellipsoid:

$$E(r) = E^\infty + \int_V (\nabla \nabla` G_{r-r`}) \, dr` \, \delta\varepsilon \, (1 - \Gamma \delta\varepsilon)^{-1} E^\infty, \qquad (A11)$$

where the dependence from r is given by the integral included here (which for points outside the ellipsoid is not constant). Expression (A10) and (A11) give the complete solution of the problem.